# Convective boundary layer sensible and latent heat flux lidar observations and towards new model parametrizations

Fabien Gibert[1], Dimitri Edouart[1], Paul Monnier[1], Julie Collignan[1], Julio Lopez[1], Claire Cénac[1]

[1] Laboratoire de Météorologie Dynamique (LMD/IPSL), École Polytechnique, Institut Polytechnique de Paris, Sorbonne Université, École normale supérieure, PSL Research University, CNRS, École des Ponts, Palaiseau, France , E-mail: gibert@lmd.polytechnique.fr

**Abstract.** Model parametrizations in the convective boundary layer are still at work especially in the interfacial layers with the surface and the free troposphere. The present paper reports simultaneous turbulence-scale lidar observations of wind speed, temperature and specific humidity in the convective boundary layer in temperate and semi-arid regions. The collected data are used to asess new parametrizations in particular in the entrainment layer.

## 1    Introduction

The turbulent transport of heat, matter and momentum in the convective boundary layer (CBL) is a key process to understand the 3D distribution of scalars in the atmosphere. Particularly critical are the surface and the entrainment layers that buffer the CBL, at the bottom with the soil and at the top with the free troposphere, respectively. If the surface layer exchanges are usually well documented by in situ observations and well understood following Monin-Obukhov similarity theory (MOST), the entrainment processes are still an unknown field both for observations and models. However, a robust parametrization of entrainment fluxes is essential for transport models to link surface fluxes and tropospheric vertical profiles of scalars such as $H2O$, $CO2$ and $CH4$ [1]. This is of particular importance for current and future space borne GHGs missions like GOSAT, OCO, MERLIN, that try to estimate surface fluxes from column integrated mixing ratio measurements [2]. To do so, new observations are needed that can provide, first, a 3-D view of the atmosphere and second, that have turbulence-scale temporal and spatial resolutions in order to investigate flux- gradient relationships and estimate higher-order moments [3]. The observations will also have to be made in different climate zones in order to study the robustness of advanced parametrizations.

In this paper we will present a preliminary analysis of the data from a 3-D mobile lidar observatory that provides simultaneous measurements of radial wind speed, temperature and specific humidity. Several scientific goals of this study includes (i) to address the issue of dissimilarity of scalar transport such as heat and water vapour in different climate regions (ii) to test model parametrization of the interfacial layer especially for fluxes and scalar variances. (iii) to assess the relevance of MOST which links gradient and flux close to the surface in heterogeneous landscape.

## 2    Instrumental deployment

A new wind, temperature, H2O and CO2 scanning lidars mobile observatory has been developed at Laboratoire de Météorologie Dynamique, Ecole Polytechnique, France, during the last years. It is detailed in paper 017. Two different scanning lidars are involved: 1) a temperature and water vapour Raman lidar at 355 nm; 2) a prototype DIAL and Doppler lidar at 2051 nm with a coherent and direct detection for radial wind speed and CO2 absorption measurements. The mobile observatory was operated in a temperate region at SIRTA observatory, Palaiseau, France and then in the semi-arid area of Lleida, Spain during the LIAISE experiment (Land surface interactions with the atmosphere over the Iberian semi-arid environment) in July 2021 (Fig. 1).

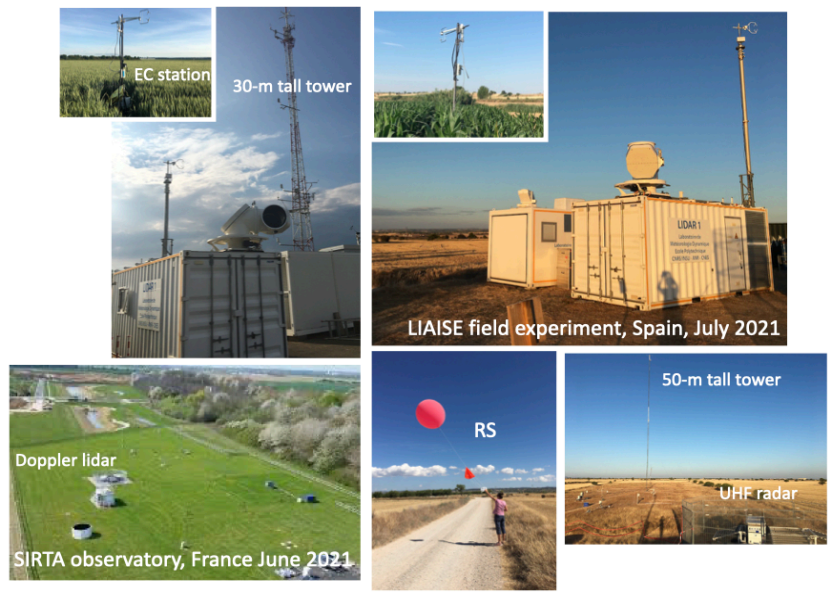

**Fig. 1.** 3-D lidar observatory and other instrumentation used in this study during two field experiments in a temperate region at SIRTA observatory, France (left panels) and in a semi-arid region with irrigated fields during LIAISE experiment, Spain (right panels). EC: eddy-covariance station, RS: radiosondes.

The LMD lidar measurements were completed by routine in situ instrumentation in the surface layer, pressure, temperature and specific humidity, wind speed and direction with colocated 30-m (SIRTA) and 50-m (LIAISE) tall tower. Daily (SIRTA) and hourly (LIAISE) soundings were also made which enables first to calibrate and validate LMD Raman lidar measurements and second to provide lidar-independent profiles for temperature, humidity and wind. Several eddy-covariance stations with sonic anemometers and gas analysers were located close to lidar observatory at different height but also at few kilometres apart to provide turbulence-linked surface layer data like sensible and latent heat fluxes, friction and free-convection scaling velocities, Monin-Obukhov length. Other remote sensing instrument like radar UHF or Doppler lidar were also used to get continuous profiles of wind speed and shear.

## 3     Lidar eddy-covariance flux

Lidar eddy-covariance measurements require the co-location of the two lidars used to provide vertical wind speed profiles for one part and temperature and specific humidity profiles on the other part. In addition, the two lidar scanning device azimuth and elevation angles are referenced using specify surface targets with known altitude with respect to lidar observatory. Even if the two lidar acquisition systems are synchronised, a cross-covariance of the data is used to correct for a potential lag time between the two time series (Fig. 2).

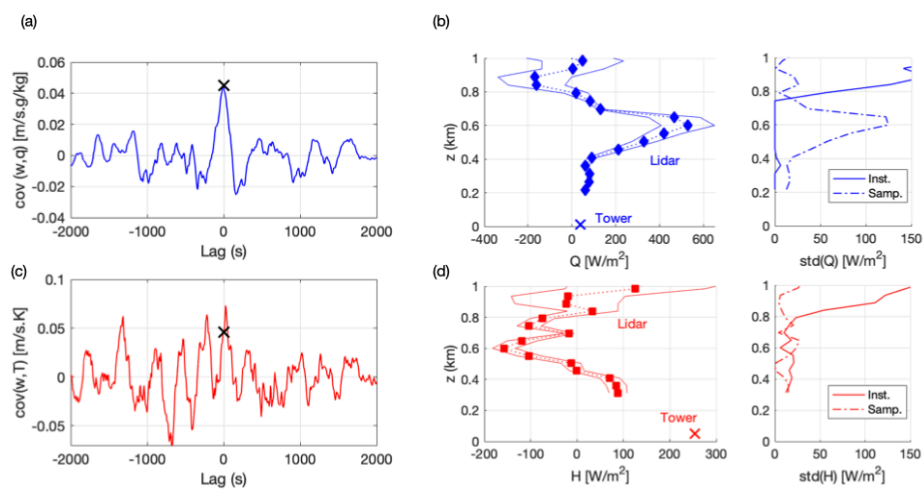

**Fig. 2.** (a) and (c) Cross-covariance between lidar vertical wind speed and specific humidity (a) and temperature (c) at a given height. (b) Latent heat flux profile and associated instrumental and sampling error. (d) Sensible heat flux profile and associated instrumental and sampling error. Time and space resolution are respectively 30 min and 50 m. Eddy-covariance in situ surface layer fluxes are indicated with an X.

The maximum is an estimate of the turbulent flux as clearly seen in Fig. 2a. However, this maximum is sometimes not clearly seen for temperature and, in that case, we use the lag position given by water vapour. Eddy-covariance flux estimates using the integral of the co-spectrum is then not applicable here, like it is for in situ data [4]. Fig. 2b and 2c show latent and sensible heat flux profiles monitored in the convective boundary layer (CBL) during LIASE experiment with the associated uncertainty that is accounted by both instrumental and sampling errors [5]. The native data used to make such calculations have a time and space resolution of 8 s and 50 m, respectively in order to respect integral scales of turbulence and avoid biaises in flux estimates.

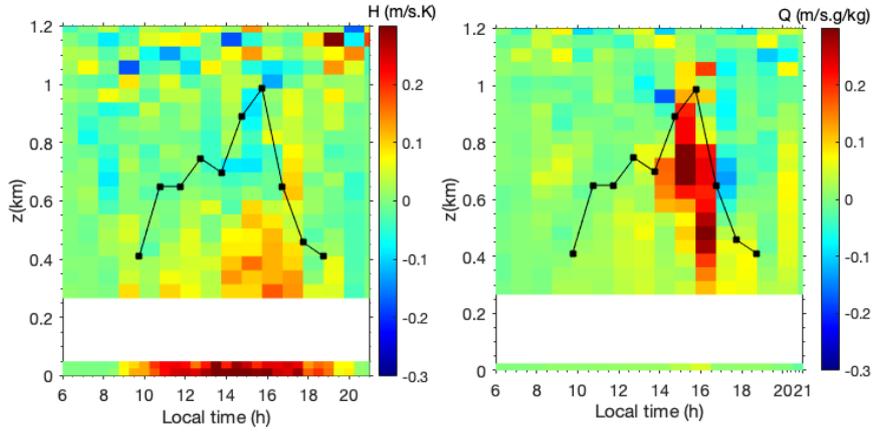

**Fig. 3.** Lidar eddy covariance sensible heat (H) and latent heat (Q) flux profiles. Black line is for the height of the convective boundary layer estimated with the minimum of sensible heat flux. Surface layer flux measurements from the 50-m tall tower are also indicated.

The flux profiles in Fig. 3 are actually typical profiles that can be measured in semiarid environment with surface large sensible and negligible latent heat fluxes [6]. Entrainment fluxes depend on surface properties as well as the strength of temperature inversion, the wind shear at the top of the CBL and water vapour gradient at the interface with the free troposphere.

## 4 First insight in gradient and flux relationships in CBL entrainment layer

### 4.1 Theoretical considerations

In a similar way of molecular diffusion one expects that the flux through an interface is related to the mean gradient of the scalar. An eddy diffusivity coefficient K for each scalar, i.e. virtual potential temperature and specific humidity in our case, may be defined as follows:

$$\overline{w'\theta_v'} = K_\theta \frac{\partial \theta_v}{\partial z} ; \overline{w'q'} = K_q \frac{\partial q}{\partial z}$$

Using a dimensional analysis and the *Buckingham Pi Theorem*, which prescribes the optimal approach to determining a dependent variable in a physical problem, Obukov built in 1946 similarity laws that linked gradient and flux in the surface layer. The same approach was used by Sorbjan et al. [7, 8] for the entrainment layer. Identifying the main physical parameters at work in this layer, i.e. surface heat flux, CBL height, interfacial gradient of temperature and specific humidity, wind shear, one may build a set of equations to link gradient and flux:

$$K_\theta(z_i) = \frac{w_*^2}{N_E} C_H f_H(Ri_E) ; K_q(z_i) = \frac{w_*^2}{N_E} C_Q f_Q(Ri_E) \, with f_{H,Q} = \frac{1 + c_{H,Q}/Ri_E}{\sqrt{1 + 1/Ri_E}}$$

where $N_E$ is the Brunt-Väisälä frequency and $Ri_E$ is the gradient Richardson number in the entrainment layer. We kept Sorbjan et al. notation with constants $C_H$ and $C_Q$ taking out of the $f_H$ and $f_Q$ functions.

## 1.1 Assessment of entrainment flux parametrization with observations

A first assessment of scalar gradient and flux link in the interfacial layer was made using the dataset collected during LIAISE experiment. The main reason is that hourly radiosondes were made at the LMD lidar site during several days which enables to check lidar data and to have two independent dataset to calculate $f_H$ and $f_Q$ functions.

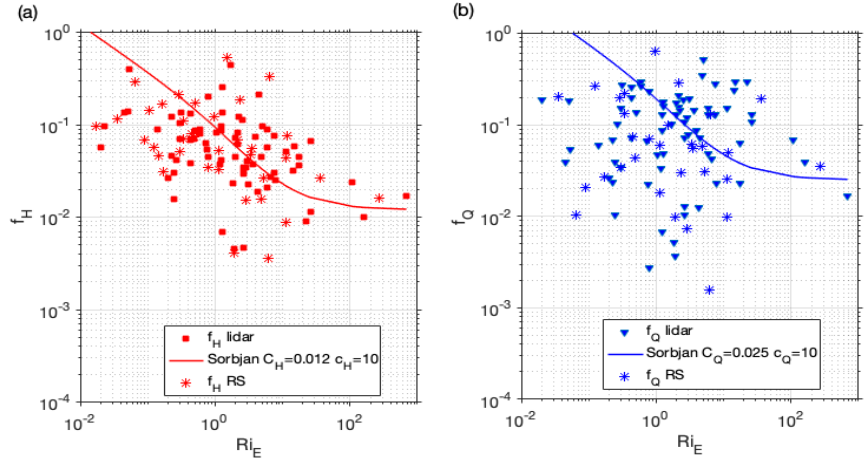

**Fig. 4.** Scaling fluxes in dependence of $Ri_E$ using lidar (square) and radiosondes (stars) dataset.

In the first dataset we estimated the scalar gradient in the entrainment layer using Raman lidar measurements at high space resolution (7.5 m) and moderate time resolution (2 min). The reason is that 50 m is a too coarse resolution to capture the gradient in sharp inversion. The wind shear and then Richardson number was calculated using a second Doppler lidar and the radar UHF. In the second dataset, radiosondes were used to calculate scalar gradient, wind shear, Brunt-Väisälä frequency and Richardson number. In the two cases, we used eddy covariance lidar estimates of fluxes in the entrainment layer. Fig. 4 shows the results for temperature and specific humidity. First comment is that C-constants suggested by Sorbjan et al. (that were found with large eddy simulations (LES) tests) can also match the observations. The evolution of the temperature data seems to follow the theoretical linear evolution (in log scale) for Richardson number larger than one but not for lower values (weak inversion, large wind shear). Second comment is that it seems that such parametrization does not work for specific humidity despite a better data precision. Error bars for flux and gradient which are not shown here are not sufficient to explain the scatter of the data. Finally, the two dataset used here seem to give the same results which also shows that (i) lidar gradient and radar wind shear estimate uncertainties are not a relevant reason for parametrization mismatch (ii) as described by Sorbjan et al., the parametrizations seem to be relevant for time averaged profiles as well as punctual profiles.

**Acknowledgements**

This work is supported by CNRS with the program HyMeX and by Agence Nationale de la Recherche with the project HILAISE. We also want to thank SIRTA observatory team and UKMO team during LIAISE experiment (Jeremy Price, Jenifer Brooke) who provides additional in situ and remote sensing data to LMD lidar observations.